\documentclass[12pt]{article}
\usepackage{epsfig}
\usepackage{axodraw}

\def\ra{\rightarrow}
\def\simlt{\stackrel{<}{{}_\sim}}
\def\simgt{\stackrel{>}{{}_\sim}}

\newcommand{\bea}{\begin{eqnarray}}
\newcommand{\eea}{\end{eqnarray}}
\newcommand{\bd}{\begin{displaymath}}
\newcommand{\ed}{\end{displaymath}}
\newcommand{\be}{\begin{equation}}
\newcommand{\ee}{\end{equation}}
\newcommand{\ord}{{\cal{O}}}
\newcommand{\tb}{\tan\beta}

\def \Oi{{\mathcal O}}

\renewcommand{\baselinestretch}{1.2}
\begin{document}

\thispagestyle{empty}

{\normalsize\sf
\rightline {hep-ph/0207241}
\rightline{TUM-HEP-471/02}
\rightline{IFT-02/27}
\vskip 3mm
\rm\rightline{July 2002}
}

\vskip 5mm

\begin{center}
  
  {\LARGE\bf Correlation between \boldmath{$\Delta M_s$} and
    \boldmath{$B^0_{s,d}\rightarrow \mu^+\mu^-$}\\ in Supersymmetry at
    Large \boldmath{$\tan\beta$} }

\vskip 10mm

{\large\bf Andrzej J.~Buras$^1$, Piotr H.~Chankowski$^2$,}\\
{\large\bf Janusz Rosiek$^{1,2}$ and {\L}ucja S{\l}awianowska$^2$} \\[5mm]

{\small $^1$ Physik Department, Technische Universit{\"a}t M{\"u}nchen,}\\
{\small D-85748 Garching, Germany}\\
{\small $^2$ Institute of Theoretical Physics, Warsaw University}\\
{\small Ho\.za 69, 00-681 Warsaw, Poland}

\end{center}
\vskip 5mm

\renewcommand{\baselinestretch}{1.1}

\begin{abstract}
  Considering the MSSM with the CKM matrix as the only source of
  flavour violation and heavy supersymmetric particles at large
  $\tan\beta$, we analyze the correlation between {\it the increase}
  of the rates of the decays $B^0_{s,d}\rightarrow \mu^+\mu^-$ and
  {\it the suppression} of $\Delta M_s$, that are caused by the
  enhanced flavour changing neutral Higgs couplings to down-type
  quarks.  We give analytic formulae for the neutral and charged Higgs
  couplings to quarks including large $\tan\beta$ resummed corrections
  in the $SU(2)\times U(1)$ limit and comment briefly on the accuracy
  of this approximation.  For $0.8\le (\Delta M_s)^{\rm exp}/(\Delta
  M_s)^{\rm SM}\le 0.95$ we find $6\cdot 10^{-7}\ge BR(B^0_s\ra
  \mu^+\mu^-)^{\rm max} \ge 4\cdot 10^{-8}$ and $1.4\cdot 10^{-8}\ge
  BR(B^0_d\ra \mu^+\mu^-)^{\rm max}\ge 1\cdot 10^{-9}$.  For $(\Delta
  M_s)^{\rm exp} \ge (\Delta M_s)^{\rm SM}$ substantial enhancements
  of $B^0_{s,d}\ra\mu^+\mu^-$ relative to the expectations based on
  the Standard Model are excluded.  With $(\Delta M_s)^{\rm
    exp}>15.0/$ps a conservative analysis of $(\Delta M_s)^{\rm SM}$
  gives $BR(B^0_s\ra \mu^+\mu^-)\simlt1.2\cdot10^{-6}$ and
  $BR(B^0_d\ra \mu^+\mu^-)\simlt3\cdot10^{-8}$.  However, we point out
  that in the less likely scenario in which the squark mixing is so
  large that the neutral Higgs contributions dominate $\Delta M_s$,
  the rates for $B^0_{s,d}\rightarrow \mu^+\mu^-$ increase with
  increasing $\Delta M_s$ and the bounds in question are weaker.
  Violation of all these correlations and bounds would indicate new
  sources of flavour violation.
\end{abstract}
\renewcommand{\baselinestretch}{1.2}

\newpage 
\setcounter{page}{1} 
\setcounter{footnote}{0}

\section{Introduction}
\setcounter{equation}{0} 

The Minimal Supersymmetric Standard Model (MSSM), with large value of
$\tan\beta$, the ratio of the two vacuum expectation values $v_u/v_d$,
is a very interesting scenario.  On the one hand, it is consistent
with the unification of the top and bottom Yukawa couplings predicted
by some $SO(10)$ GUT models.  On the other hand, its predictions for
rates of certain low energy processes can differ significantly from
the ones of the Standard Model (SM) even for heavy sparticles and with
the Cabibbo-Kobayashi-Maskawa (CKM) matrix being the only source of
flavour and CP violation in the quark sector.

In the down-quark sector large supersymmetric effects originate from
$\tan\beta$ enhanced flavour changing neutral currents (FCNC) mediated
by Higgs scalars and generated at one loop by Higgs penguin-like
diagrams with charginos and top-squarks. They have been first
considered in \cite{HAPOTO} and subsequently found to {\it increase}
by orders of magnitude the branching ratios of the rare decays
$B^0_{s,d}\to\mu^+\mu^-$~\cite{BAKO,CHSL,BOEWKRUR,HULIYAZH} and to
{\it decrease} significantly the $B^0_s$-$\bar B^0_s$ mass difference
$\Delta M_s$~\cite{BUCHROSL} relative to the expectations based on the
SM.  Since both these effects are caused by the same neutral Higgs
boson mediated FCNC (see figs.~\ref{fig:2pg} and~\ref{fig:bmumu}), a
correlation between them must exist~\cite{BUCHROSL}.  This is
particularly interesting as $\Delta M_s$ and
$BR(B^0_{s,d}\to\mu^+\mu^-)$ can in principle be measured at the
Tevatron and $B$--factories in the coming years.  It is the purpose of
this letter to point out the consequences of this correlation.

Analyzing low energy processes in the MSSM with $\tan\beta\gg1$ it is
essential to take into account all potentially large effects in a
consistent framework.  Four such effects have been identified in the
literature:

1) Modification of the tree-level relation between the MSSM Lagrangian
mass parameters $m_d$, $m_s$, $m_b$ determining the corresponding
Yukawa couplings and the running (``measured'') quark masses
${\overline m_d}$, ${\overline m_s}$, ${\overline m_b}$~\cite{HARASA}.

2) Corrections to the CKM matrix, as a result of which elements of the
physical CKM matrix, to be called $V_{JI}^{\rm eff}$, differ from
$V_{JI}$ present in the original Lagrangian~\cite{BLPORA}.

3) Enhanced flavour changing neutral Higgs boson penguins mentioned
above.

4) Enhanced corrections to charged Higgs boson vertices~\cite{DEGAGI}.

Several steps towards including consistently all these effects in
phenomenological analyses have been already made during the last
years.  In refs.~\cite{DEGAGI,CAGANIWA1,CAGANIWA2} the effects 1) and
4) have been discussed in the context of the $\bar B\to X_s\gamma$
decay.  In~\cite{ISRE} the effects 1)-3) have been calculated in the
$SU(2)\times U(1)$ symmetry limit in the context of
$B^0_{s,d}\to\mu^+\mu^-$ decays and $B^0_{s,d}$-$\bar B^0_{s,d}$
mixings confirming the increase of $BR(B^0_{s,d}\to\mu^+\mu^-)$ and
the suppression of $\Delta M_s$ pointed out
in~\cite{BAKO,CHSL,BOEWKRUR,HULIYAZH} and~\cite{BUCHROSL},
respectively.

In the following detailed analysis~\cite{BUCHROSL02} we extend these
analyses based on $SU(2)\times U(1)$ symmetry limit~\cite{BAKO,ISRE}
by calculating all the four effects in a more general effective
Lagrangian approach, comparing the results, analytically and
numerically, with the $SU(2)\times U(1)$ symmetry limit and thereby
confirming and in certain cases correcting and generalizing analytical
rules for inclusion of the large $\tb$ effects presented
in~\cite{DEGAGI,CAGANIWA1,CAGANIWA2,ISRE}.  As the analysis
of~\cite{BUCHROSL02} is long and technical, in the present letter we
summarize compactly the results for all the four listed effects.  We
present numerical results based on the formalism of~\cite{BUCHROSL02}
and explain them qualitatively using the formulae obtained in the
$SU(2)\times U(1)$ symmetry limit.  This allows us to analyze in
detail the correlation between $BR(B^0_{s,d}\to\mu^+\mu^-)$ and
$\Delta M_s$ pointed out in~\cite{BUCHROSL} taking into account the
$\bar B\rightarrow X_s\gamma$ constraint.

During the completion of this letter a model independent analysis of
rare processes in theories with the CKM matrix as the unique source of
flavour and CP violation has been presented in~\cite{AMGIISST}.  While
those authors also investigated large $\tan\beta$ effects in
$BR(B^0_{s,d}\to\mu^+\mu^-)$, $\Delta M_s$ and $\bar B\rightarrow
X_s\gamma$, they have not analyzed the correlation between
$BR(B^0_{s,d}\to\mu^+\mu^-)$ and $\Delta M_s$ addressed here.

As the recent discussions in the literature~\cite{BOEWKRUR,AMGIISST}
show that the statements like ``models in which flavour mixing is
ruled only by the CKM matrix'' or ``models with minimal flavour
violation'' have different meaning in different papers, we would like
to specify the structure of flavour violation in the MSSM version
considered by us.  While the flavour violation in the scenario
considered is ruled by the CKM matrix, it should be emphasized that
for split soft SUSY breaking masses of left-handed squarks belonging
to different generations some flavour violation unavoidably appears in
the up- or down-type (or in both) squark mass squared matrices.  In
our calculations we choose the soft SUSY breaking mass parameter
$m^2_Q$ such that flavour violation appears in the up-type squark mass
matrix.  The scenario with flavour violation in the down-type squark
mass matrix would require the inclusion of box and Higgs penguin
diagrams with gluinos and is beyond the scope of this paper.

\section{The effective Lagrangian}
\setcounter{equation}{0}

Let us consider the decoupling of sparticles in the limit of unbroken
$SU(2)\times U(1)$ symmetry~\cite{BAKO,ISRE}.  The electroweak
symmetry breaking is then taken into account after sparticles are
integrated out.  This approximation should be valid if the sparticle
mass scale is larger than that of the Higgs boson sector (set by
$M_{H^+}$). The absence of vacuum expectation values before decoupling
implies neglecting the left-right mixing in the squark mass matrices
even for non-vanishing $A_{u,d}$ and/or $\mu$ parameters.

In this approach below the sparticle mass scale the effective
Lagrangians describing the neutral and charged Higgs boson couplings
to the down- and up-type quarks have the form~\cite{CHPO,BAKO}
\begin{eqnarray}
{\cal L}_{\rm eff}^{(d)} =-\epsilon_{ij}H^{(d)}_i\overline{d_R}\cdot(
\mathbf{Y}_d+\Delta_d\mathbf{Y}_d)\cdot q_{jL}
-H^{(u)\ast}_i\overline{d_R}\cdot \Delta_u\mathbf{Y}_d\cdot q_{iL}
+{\rm H.c.}\label{eqn:effL_1}
\end{eqnarray}
\begin{eqnarray}
{\cal L}_{\rm eff}^{(u)} = -\epsilon_{ij}H^{(u)}_i\overline{u_R}\cdot
\left(\mathbf{Y}_u+\Delta_u\mathbf{Y}_u\right)\cdot q_{jL}
-H^{(d)\ast}_i\overline{u_R}\cdot \Delta_d\mathbf{Y}_u\cdot q_{iL}
+{\rm H.c.}\label{eqn:effL_2}
\end{eqnarray}
where $\epsilon_{21}=-\epsilon_{12}=1$ and $\mathbf{Y}_{d,u}$ are
Yukawa coupling matrices.  The neutral and charged components of the
two Higgs doublets are given in the standard way
\begin{eqnarray}\label{DECOMP}
H^{(d)}_1={v_d\over\sqrt2}+{1\over\sqrt2}\left(H^0\cos\alpha
-h^0\sin\alpha +iA^0\sin\beta -iG^0\cos\beta \right)\nonumber \\
H^{(u)\ast}_2={v_u\over\sqrt2}+ {1\over\sqrt2}\left(H^0\sin\alpha +h^0
\cos\alpha -iA^0 \cos\beta -iG^0 \sin\beta \right)
\end{eqnarray}
\begin{eqnarray}
H^{(d)*}_2 = H^+\sin\beta - G^+\cos\beta, \qquad
H^{(u)}_1 = H^+\cos\beta + G^+\sin\beta~.
\label{PHYS}
\end{eqnarray}
In these conventions 
\be\label{mass}
m_{d_J}=-{v_d\over\sqrt2} y_{d_J}, \qquad
m_{u_J}= {v_u\over\sqrt2} y_{u_J}
\ee
where $y_{d_J}$ and $y_{u_J}$ are the Yukawa couplings.  Here $J$ is
the flavour index with $d_1\equiv d$, $d_2\equiv s$, $d_3\equiv b$ and
similarly for the up-type quarks.  Finally $v_d^2/\cos^2\beta =
v_u^2/\sin^2\beta=1/\sqrt2G_F\approx (246~{\rm GeV})^2$.

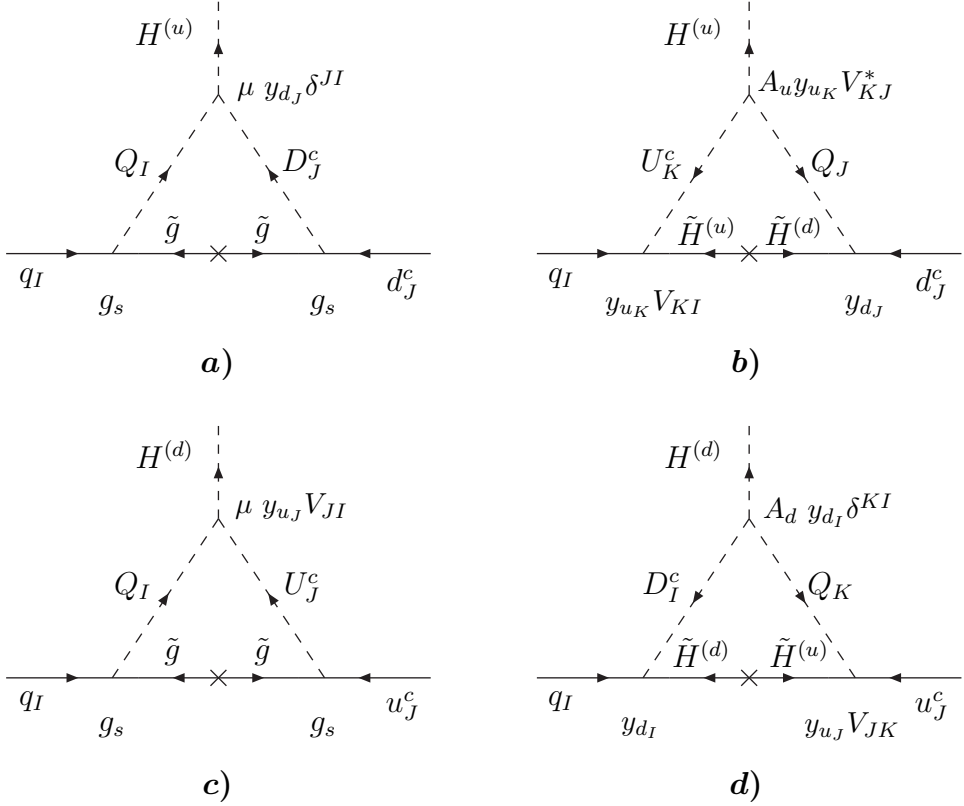
\begin{figure}[t] 
\begin{center}
\begin{picture}(400,320)(0,0)
\ArrowLine(90,200)(120,200)
\ArrowLine(90,200)(60,200)
\ArrowLine(10,200)(60,200)
\ArrowLine(170,200)(120,200)
\Line(87,203)(93,197)
\Line(87,197)(93,203)
\Text(20,195)[t]{$q_I$}
\Text(160,195)[t]{$d^c_J$}
\Text(73,215)[t]{$\tilde{g}$}
\Text(107,215)[t]{$\tilde{g}$}
\Text(58,240)[t]{$Q_I$}
\Text(122,240)[t]{$D^c_J$}
\Text(70,290)[t]{$H^{(u)}$}
\Text(50,185)[t]{$g_s$}
\Text(130,185)[t]{$g_s$}
\Text(118,270)[t]{\small $\mu ~y_{d_J}\delta^{JI}$}
\DashArrowLine(50,200)(90,260){4}
\DashArrowLine(130,200)(90,260){4}
\DashArrowLine(90,260)(90,295){4}
\Text(90,165)[t]{$\mbox{\boldmath$a)$}$}
\ArrowLine(290,200)(320,200)
\ArrowLine(290,200)(260,200)
\ArrowLine(210,200)(260,200)
\ArrowLine(370,200)(320,200)
\Line(287,203)(293,197)
\Line(287,197)(293,203)
\Text(220,195)[t]{$q_I$}
\Text(360,195)[t]{$d^c_J$}
\Text(275,215)[t]{$\tilde H^{(u)}$}
\Text(308,215)[t]{$\tilde H^{(d)}$}
\Text(258,240)[t]{$U^c_K$}
\Text(322,240)[t]{$Q_J$}
\Text(270,290)[t]{$H^{(u)}$}
\Text(255,187)[t]{\small $y_{u_K}V_{KI}$}
\Text(335,185)[t]{$y_{d_J}$}
\Text(320,270)[t]{\small $A_uy_{u_K}V^\ast_{KJ}$}
\DashArrowLine(290,260)(250,200){4}
\DashArrowLine(290,260)(330,200){4}
\DashArrowLine(290,260)(290,295){4}
\Text(290,165)[t]{$\mbox{\boldmath$b)$}$}
\ArrowLine(90,40)(120,40)
\ArrowLine(90,40)(60,40)
\ArrowLine(10,40)(60,40)
\ArrowLine(170,40)(120,40)
\Line(87,43)(93,37)
\Line(87,37)(93,43)
\Text(20,35)[t]{$q_I$}
\Text(160,35)[t]{$u^c_J$}
\Text(50,25)[t]{$g_s$}
\Text(130,25)[t]{$g_s$}
\Text(73,55)[t]{$\tilde{g}$}
\Text(107,55)[t]{$\tilde{g}$}
\Text(58,80)[t]{$Q_I$}
\Text(122,80)[t]{$U^c_J$}
\Text(118,110)[t]{\small $\mu ~y_{u_J}V_{JI}$}
\Text(70,130)[t]{$H^{(d)}$}
\DashArrowLine(50,40)(90,100){4}
\DashArrowLine(130,40)(90,100){4}
\DashArrowLine(90,100)(90,135){4}
\Text(90,5)[t]{$\mbox{\boldmath$c)$}$}
\ArrowLine(290,40)(260,40)
\ArrowLine(290,40)(320,40)
\ArrowLine(210,40)(260,40)
\ArrowLine(370,40)(320,40)
\Line(287,43)(293,37)
\Line(287,37)(293,43)
\Text(220,35)[t]{$q_I$}
\Text(360,35)[t]{$u^c_J$}
\Text(273,55)[t]{$\tilde H^{(d)}$}
\Text(310,55)[t]{$\tilde H^{(u)}$}
\Text(258,80)[t]{$D^c_I$}
\Text(322,80)[t]{$Q_K$}
\Text(321,110)[t]{\small $A_d ~y_{d_I}\delta^{KI}$}
\Text(270,130)[t]{$H^{(d)}$}
\Text(250,25)[t]{\small $y_{d_I}$}
\Text(330,27)[t]{\small $y_{u_J}V_{JK}$}
\DashArrowLine(290,100)(250,40){4}
\DashArrowLine(290,100)(330,40){4}
\DashArrowLine(290,100)(290,135){4}
\Text(290,5)[t]{$\mbox{\boldmath$d)$}$}
\end{picture}
\end{center}
\vspace{-5mm}
\caption{Vertex corrections in the $SU(2)\times U(1)$ symmetry limit.   
  Diagrams a) and b) give rise to corrections
  $(\Delta_u\mathbf{Y}_d)^{JI}$, diagrams c) and d) to corrections
  $(\Delta_d\mathbf{Y}_u)^{JI}$.  }
\label{fig:bcrs1}
\end{figure}

The loop induced terms $\Delta_d\mathbf{Y}_d$ and
$\Delta_u\mathbf{Y}_u$ are always subleading in the large $\tan\beta$
limit and can be neglected.  Diagrams giving rise to the correction
$\Delta_u\mathbf{Y}_d$ are shown in figs.~\ref{fig:bcrs1}a
and~\ref{fig:bcrs1}b.  In the basis in which $\mathbf{Y}_d={\rm
  diag}(y_d)$, $\mathbf{Y}_u={\rm diag}(y_u)\cdot V$ where $V$ is the
CKM matrix, and neglecting $y^2_u$ and $y_c^2$, the correction
$\Delta_u\mathbf{Y}_d$ has the structure~\cite{ISRE}
\begin{eqnarray}
\left(\Delta_u\mathbf{Y}_d\right)^{JI}=-y_{d_J}
\left(\epsilon_0\delta^{JI}+\epsilon_Y y^2_t V^{3J\ast} V^{3I}\right).  
\label{eqn:DuYd_str}
\end{eqnarray}
The correction $\Delta_d\mathbf{Y}_u$ is generated by the diagrams
shown in figs.~\ref{fig:bcrs1}c and~\ref{fig:bcrs1}d and has the form
\be
\left(\Delta_d\mathbf{Y}_u\right)^{JI}=y_{u_J}V_{JI}
\left(\epsilon^\prime_0+\epsilon^\prime_Yy^2_{d_I}\right)~.
\label{eqn:DdYu_str}
\ee
The four quantities $\epsilon_0$, $\epsilon_Y$, $\epsilon^\prime_0$,
$\epsilon^\prime_Y$ can be obtained by calculating the diagrams in
fig.~\ref{fig:bcrs1}:
\begin{eqnarray}
\epsilon_0=-{2\alpha_s\over3\pi}{\mu\over m_{\tilde g}}
H_2\left(x^{Q/g},x^{D/g}\right), \phantom{aaaa}
\epsilon_Y={1\over16\pi^2} \frac{A_t}{\mu}~
H_2\left(x^{Q/\mu},x^{U/\mu}\right)\label{eqn:e}\\
\epsilon^\prime_0=-{2\alpha_s\over3\pi}{\mu\over m_{\tilde g}}
H_2\left(x^{Q/g},x^{U/g}\right), \phantom{aaaa}
\epsilon^\prime_Y={1\over16\pi^2} 
\frac{A_b}{\mu}~ H_2\left(x^{Q/\mu},x^{D/\mu}\right)
\label{eqn:eprime}
\end{eqnarray}
where $x^{Q/g}\equiv m^2_Q/m^2_{\tilde g}$, $x^{D/g}\equiv
m^2_D/m^2_{\tilde g}$, $x^{Q/\mu}\equiv m^2_Q/\mu^2$ etc., and
$m^2_Q$, $m^2_D$, $m^2_U$, $A_t$, and $A_b$ are the parameters of the
soft supersymmetry breaking in the MSSM Lagrangian\footnote{Our
  convention~\cite{ROS} for $A_u$ and $A_d$ parameters is fixed by the
  form of the left-right mixing terms in the squark mass matrices
  which read $-m_u(A_u+\mu\cot\beta)$ and $-m_d(A_d+\mu\tan\beta)$ for
  the up and down squarks, respectively.}.  The function $H_2(x,y)$ is
defined as
\begin{eqnarray}
H_2(x,y) = {x\ln x\over(1-x)(x-y)}+{y\ln y\over(1-y)(y-x)}~.
\label{eqn:hfun}
\end{eqnarray}
The eqs. (10),(15),(16) of~\cite{DEGAGI} reduce to~(\ref{eqn:e})
and~(\ref{eqn:eprime}) in the $SU(2)\times U(1)$ symmetry limit.

\section{Effective Parameters and Couplings}
\setcounter{equation}{0}

The mass matrices of the down- and up-type quarks can be obtained by
replacing the neutral scalar fields in~(\ref{eqn:effL_1})
and~(\ref{eqn:effL_2}) by their vacuum expectation values.  One finds
that the down-type-quark mass matrix $\hat M_d$ receives $\tan\beta$
enhanced corrections both to the diagonal and non-diagonal entries,
whereas the corresponding corrections to $\hat M_u$ are negligible.
$\hat M_d$ is then diagonalized by the appropriate rotations of the
$d_L$ and $d_R$ fields.  Except for the charged Higgs boson $H^+$
couplings in which loop correction $\Delta_d\mathbf{Y}_u$ matters, the
four effects listed in the Introduction result from performing these
rotations on the $d_L$ and $d_R$ fields in the interaction vertices
in~(\ref{eqn:effL_1}) and~(\ref{eqn:effL_2}).

In the full approach that goes beyond the $SU(2)\times U(1)$ symmetry
limit~\cite{BUCHROSL02}, the corrections to $\hat M_d$ are found by
calculating directly the self-energy diagrams of the down-type-quarks.
The resulting formulae are rather complicated and are presented
in~\cite{BUCHROSL02} where also the derivation of the formulae in the
$SU(2)\times U(1)$ limit is described in detail.

Below we give the formulae that summarize the effects 1)--4) in the
$SU(2)\times U(1)$ symmetry limit.  The quark fields in these formulae
are mass eigenstates of the one-loop corrected matrices $\hat M_d$ and
$\hat M_u$ as opposed to the original fields in~(\ref{eqn:effL_1})
and~(\ref{eqn:effL_2}).

{\bf 1.} The original mass parameters $m_{d_J}$ and $m_{u_J}$
in~(\ref{mass}), that enter the Feynman rules, are related to the {\it
  effective} running mass parameters $\overline m_{d_J}$ and
$\overline m_{u_J}$ of the low energy theory through~\cite{HARASA}
\be\label{basic}
m_{d_J}=\frac{\overline m_{d_J}}{1+\tilde\epsilon_J\tan\beta},
\qquad  
m_{u_J}\approx \overline m_{u_J}
\ee
with $\tilde\epsilon_J$ given by 
\be
\label{epJ}
\tilde\epsilon_J\equiv\epsilon_0+\epsilon_Yy^2_tV_{3J}^\ast V_{3J}
\approx \epsilon_0+\epsilon_Yy^2_t\delta^{J3}~.
\ee
It has been shown~\cite{CAGANIWA1} that expressing $m_{d_J}$ through
${\overline m_{d_J}}$ by means of~(\ref{basic}) in the neutral and
charged Higgs couplings resums for large values of $\tb$ dominant
supersymmetric corrections to all orders of perturbation theory.  Such
a resummation is necessary for obtaining reliable results.  Note that
in contrast to the corrections to $m_b$ in~(\ref{basic}), the ones to
$m_d$ and $m_s$ do not depend on the top Yukawa coupling.
  
{\bf 2.} The original elements of the CKM matrix, $V_{JI}$, present in
the Feynman rules of the MSSM are related to the {\it effective} CKM
matrix $V_{JI}^{\rm eff}$ through~\cite{BLPORA,BAKO,ISRE,BUCHROSL02}
\begin{eqnarray}
&&V_{JI}=V_{JI}^{\rm eff} ~
\left[{1+\tilde\epsilon_3\tan\beta\over1+\epsilon_0\tan\beta}\right]
\phantom{aaa}{\rm for}\phantom{aa} 
(JI)=(13), ~(23), ~(31) ~{\rm and} ~(32),\nonumber\\
&&V_{JI}=V_{JI}^{\rm eff}\phantom{aaa}{\rm otherwise.}
\label{eqn:CKMcorr}
\end{eqnarray}
It is $V_{JI}^{\rm eff}$ that has to be identified with the CKM matrix
whose elements are determined from the low energy processes.  Note
that the elements $|V_{ub}|$ and $|V_{cb}|$, that are affected by
these corrections are usually determined from tree level decays under
the assumption that new physics contributions to the relevant
branching ratios can be neglected.  This assumption is violated in the
case of supersymmetry at large $\tan\beta$.  In other words, what
experimentalists extract from tree level decays are $|V^{\rm
  eff}_{ub}|$ and $|V^{\rm eff}_{cb}|$ and not $|V_{ub}|$ and
$|V_{cb}|$.

{\bf 3.} The effective Lagrangian describing flavour violating neutral
Higgs interactions with down-type quarks is given by
\begin{eqnarray}\label{LNEUTRAL}
{\cal L}^{\rm off-diag}_{\rm eff}=
-\overline{(d_J)_R}\left[X^S_{RL}\right]^{JI}(d_I)_L S^0
-\overline{(d_J)_L}\left[X^S_{LR}\right]^{JI}(d_I)_R S^0
\end{eqnarray}
with $S^0=(H^0,h^0,A^0,G^0)$.  In the case of $B$-physics the pairs
$(J,I)=(3,2),(3,1),(2,3)$ and $(1,3)$ matter.  We
find~\cite{BUCHROSL02}
\be 
\left[X^S_{RL}\right]^{JI} =\left[X^S_{LR}\right]^{IJ*}=\frac{g}{2
  M_W\cos\beta} \frac{{\overline m}_{d_J} V^{3J*}_{\rm
    eff}V^{3I}_{\rm eff}} {(1 + \tilde\epsilon_3\tan\beta)(1 +
  \epsilon_0\tan\beta)} \epsilon_Yy^2_t \left(x_u^S -
  x^S_d\tan\beta\right)
\label{BXRLFIN}
\ee
where $x^S_d=(\cos\alpha,-\sin\alpha,i\sin\beta,-i\cos\beta)$, and
$x^S_u=(\sin\alpha,\cos\alpha,-i\cos\beta,-i\sin\beta)$.

In the case of $K$-physics the pairs $(J,I)=(2,1)$ and $(1,2)$ matter
and we find~\cite{BUCHROSL02}
\be
\left[X^S_{RL}\right]^{JI}=\left[X^S_{LR}\right]^{IJ*}
=\frac{g}{2 M_W\cos\beta} {\overline m}_{d_J} V^{3J*}_{\rm
  eff}V^{3I}_{\rm eff} \frac{(1 + \tilde\epsilon_3\tan\beta)^2}{(1
  + \epsilon_0\tan\beta)^4} \epsilon_Yy^2_t
\left(x_u^S-x^S_d\tan\beta\right)~.
\label{KXRLFIN}
\ee
Note that the flavour violating couplings of $G^0$ vanish in this
limit.  Formulae~(\ref{BXRLFIN})--(\ref{KXRLFIN}) agree with the
recent corrected version of~\cite{ISRE} except that $V_{J3}^{{\rm
    eff}\ast}$ in equation (10) of that paper should be replaced by
$V_{3J}^{{\rm eff}\ast}$.

{\bf 4.} The effective couplings of the charged Higgs ($H^\pm$) and
Goldstone ($G^\pm$) bosons to quarks are given respectively by
\be\label{LH}
{\cal L}^{H^+}_{\rm eff}=
\overline{(u_J)_R}\left[P^H_{RL}\right]^{JI}(d_I)_L H^+ +
\overline{(u_J)_L}\left[P^H_{LR}\right]^{JI}(d_I)_R H^+ +h.c
\ee
\be\label{LG}
{\cal L}^{G^+}_{\rm eff}=
\overline{(u_J)_R}\left[P^G_{RL}\right]^{JI}(d_I)_L G^+ +
\overline{(u_J)_L}\left[P^G_{LR}\right]^{JI}(d_I)_R G^+ + h.c~.
\ee
It is useful to define the parameters $\epsilon^{HL}_{JI}$,
$\epsilon^{HR}_{JI}$, $\epsilon^{GL}_{JI}$ and $\epsilon^{GR}_{JI}$
through
\be
\left[P^H_{RL}\right]^{JI}={g\over\sqrt2M_W}\cot\beta \overline
m_{u_J} V^{\rm eff}_{JI} (1-\epsilon^{HL}_{JI}), \qquad
\left[P^H_{LR}\right]^{JI}={g\over\sqrt2M_W}\tan\beta V^{\rm
  eff}_{JI} \overline m_{d_I} (1-\epsilon^{HR}_{JI}),\nonumber
\ee
\be
\left[P^G_{RL}\right]^{JI}={g\over\sqrt2M_W} \overline m_{u_J} V^{\rm
  eff}_{JI} (1+\epsilon^{GL}_{JI}), \qquad
\left[P^G_{LR}\right]^{JI}=-{g\over\sqrt2M_W} V^{\rm eff}_{JI}
\overline m_{d_I} (1+\epsilon^{GR}_{JI}).  \nonumber
\ee
Using $\tilde\epsilon_J$ defined in~(\ref{epJ}), we find in the
$SU(2)\times U(1)$ symmetry limit~\cite{BUCHROSL02}
\begin{eqnarray}
\epsilon^{HL}_{JI}=
\tan\beta\left(\epsilon_0^\prime+\epsilon_Y^\prime y^2_b \delta^{I3}\right)
+\Delta_{JI}~
\qquad
\epsilon^{HR}_{JI}=
\frac{\tilde\epsilon_J\tan\beta}{1+\tilde\epsilon_J\tan\beta}~,
\phantom{aaa} \epsilon^{GL}_{JI}=\epsilon^{GR}_{JI}=0.
\label{F1}
\end{eqnarray}
where
\be\label{delta}
\Delta_{JI}=y^2_by^2_t \frac{\epsilon_Y\epsilon_Y^\prime \tan^2\beta}
{1+\epsilon_0\tan\beta}
\times\left\{ \begin{array}{ll}
+1 &{(J,I)=(1,3),(2,3)}\\
-1 & {(J,I)=(3,1),(3,2)} \\
0  &{\rm otherwise}
\end{array} \right.
\ee
In the $SU(2)\times U(1)$ symmetry limit vanishing of the corrections 
$\epsilon^{GL}_{JI}$ and $\epsilon^{GR}_{JI}$ to the charged Goldstone 
boson vertices expressed in terms of $V^{\rm eff}$ and physical masses 
$\overline m_{d_I}$ is required by gauge invariance~\cite{BUCHROSL02}. The 
results for $\epsilon^{HL}_{JI}$ and $\epsilon^{HR}_{JI}$ agree with ref. 
\cite{AMGIISST}, where the presence of $\Delta_{JI}$ has been pointed out.

We observe that the $\epsilon^{HR}_{JI}$ corrections to the vertices
involving $V^{\rm eff}_{ts}$ and $V^{\rm eff}_{td}$ depend on the top
Yukawa coupling $y_t^2$ while those to the vertices involving $V^{\rm
  eff}_{cb}$ and $V^{\rm eff}_{ub}$ do not. Note also that whereas the
rule~(\ref{F1}) for $\epsilon^{HR}_{JI}$ for $(J\not=3,I)$ and
accidentally for $J=I=3$ is equivalent to expressing in the tree level
formulae $m_{d_I}$ and $V_{JI}$ through $\overline m_{d_I}$ and
$V^{\rm eff}_{JI}$ by means of~(\ref{basic}) and~(\ref{eqn:CKMcorr})
respectively, for $J=3$ and $I=1,2$ it is more involved.  Expressing
in these cases only $V_{JI}$ and $m_{d_I}$ through $V^{\rm eff}_{JI}$
and $\overline m_{d_I}$, would give wrong results.  In~\cite{DEGAGI}
explicit expressions for $[P^{H(G)}_{RL}]^{JI}$ with $J=3$, $I=1,2,3$
and for $[P^{H(G)}_{LR}]^{JI}$ with $J=1,2,3$ and $I=3$ have been
given omitting the modifications of the CKM factors summarized
in~(\ref{eqn:CKMcorr}) - see the formula (17) of that paper.  As
discussed in~\cite{BUCHROSL02}, the particular couplings given
in~\cite{DEGAGI} agree with the formulae given above provided
$\Delta_{JI}$ is set to zero and the CKM matrix $V$ of~\cite{DEGAGI}
is identified with $V^{\rm eff}$ in $[P^{H(G)}_{RL}]^{JI}$ of that
paper and with the original MSSM CKM matrix in $[P^{H(G)}_{LR}]^{JI}$.
In spite of this inconsistency, in the special case of the dominant
operator in the $\bar B\to X_s\gamma$ decay, the recipes for the
inclusion of large $\tan\beta$ effects into Wilson coefficients
formulated in eqs.  (18) and (19) of that paper are accidentally
correct provided all the CKM factors involved in this decay are
identified with $V_{\rm eff}$ and $\Delta_{JI}$ is set to zero.
However, as emphasized in \cite{AMGIISST} $\Delta_{JI}$ cannot be
generally neglected for $|\epsilon_Y\tan\beta|$ and
$|\epsilon_Y^\prime\tan\beta|$ larger than $0.5$ and it could be
important for $\epsilon_0^\prime\approx -\epsilon_Y^\prime$ when the
$\ord(\tan\beta)$ term in $\epsilon^{HL}_{JI}$ is small.

As discussed in detail in~\cite{BUCHROSL02}, the approximations
described here work rather well for the relation~(\ref{basic}) between
the original mass parameters $m_{d_I}$ (i.e. the Yukawa couplings) and
the running masses $\overline{m}_{d_I}$ and also for the relation
between $V$ and $V_{\rm eff}$.  The differences between the full and
approximate calculation are usually smaller than 15\% and are mainly
due to neglecting in the $SU(2)\times U(1)$ symmetry limit some gauge
coupling-dependent terms.  The same remains true also for the flavour
changing couplings $X_{RL}$ and $X_{LR}$ of the neutral scalars since
their dominant parts originate from the rotations of $d_L$ and $d_R$
fields which are directly related to the corrections to the down-type
quark mass matrix.

Let us record that typically $|\epsilon_0|$ and $|\tilde\epsilon_3|$
are $\sim 5 \times 10^{-3}$ and can reach $\sim 10^{-2}$ for very
large values of $|\mu|$ and/or $|A_t|$.  We have also checked that
taking the $\bar B\rightarrow X_s\gamma$ constraint into account,
values of the factor $(1+\tilde\epsilon_3\tan\beta)
(1+\epsilon_0\tan\beta)$ entering the denominator of
eq.~(\ref{BXRLFIN}), vary between $0.2$ and~$2$ for $\tan\beta\approx
50$.

In the case of charged Higgs boson couplings the full calculation
confirms the smallness of the corrections $\epsilon^{GL(R)}$
(typically $|\epsilon^{GL(R)}|\simlt0.05$).  The approximate
formulae~(\ref{F1}) for $\epsilon^{HR}$ and especially for
$\epsilon^{HL}$ are not as accurate as the ones for the couplings
$X_{RL}$ and $X_{LR}$.  This is because triangle vertex diagrams with
the chargino-neutralino pairs coupling to $H^+$ also play a role.
However, in the case of the $B^0_s$-$\bar B^0_s$ mixing and of the
decays $B^0_{s,d}\rightarrow\mu^+\mu^-$ these corrections constitute
only subdominant contribution to the relevant amplitudes and the
inaccuracy of the approximation is not essential.  Therefore, the
approximate formulae we present in the following section give
qualitatively correct picture of the dependence of the dominant
corrections to the $B^0_s$-$\bar B^0_s$ and $B^0_{s,d} \rightarrow
\mu^+\mu^-$ amplitudes on the MSSM parameters.  We stress however,
that the results presented in fig.~\ref{fig:bcorr} are based on the
complete calculation along the lines of~\cite{BUCHROSL02}.

\section{\boldmath{$\Delta M_s$} and  \boldmath{$B^0_{s,d}\rightarrow 
    \mu^+\mu^-$}}

\setcounter{equation}{0}

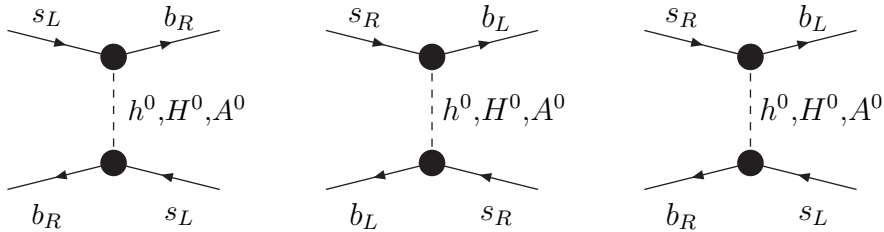
\begin{figure}[htbp]
\begin{center}
\begin{picture}(340,80)(0,0)
\ArrowLine(50,20)(10,10)
\ArrowLine(90,10)(50,20)
\Vertex(50,20){5}
\ArrowLine(10,70)(50,60)
\ArrowLine(50,60)(90,70)
\Vertex(50,60){5}
\DashLine(50,20)(50,60){3}
\Text(78,40)[]{\small $h^0$,$H^0$,$A^0$}
\Text(25,0)[]{\small $b_R$}
\Text(75,0)[]{\small $s_L$}
\Text(75,75)[]{\small $b_R$}
\Text(25,75)[]{\small $s_L$}
\ArrowLine(170,20)(130,10)
\ArrowLine(210,10)(170,20)
\Vertex(170,20){5}
\ArrowLine(130,70)(170,60)
\ArrowLine(170,60)(210,70)
\Vertex(170,60){5}
\DashLine(170,20)(170,60){3}
\Text(198,40)[]{$h^0$,$H^0$,$A^0$}
\Text(145,0)[]{$b_L$}
\Text(195,0)[]{$s_R$}
\Text(195,75)[]{$b_L$}
\Text(145,75)[]{$s_R$}
\ArrowLine(290,20)(250,10)
\ArrowLine(330,10)(290,20)
\Vertex(290,20){5}
\ArrowLine(250,70)(290,60)
\ArrowLine(290,60)(330,70)
\Vertex(290,60){5}
\DashLine(290,20)(290,60){3}
\Text(318,40)[]{$h^0$,$H^0$,$A^0$}
\Text(265,0)[]{$b_R$}
\Text(315,0)[]{$s_L$}
\Text(315,75)[]{$b_L$}
\Text(265,75)[]{$s_R$}
\end{picture}
\end{center}
\caption{Double penguin diagrams contributing to $\Delta M_s$.}
\label{fig:2pg}
\end{figure}

{\bf 1.} In the supersymmetric scenario considered here, $\Delta
M_{s}$ is given by
\be\label{DMS0}
\Delta M_s=|(\Delta M_s)^{\rm SM}+(\Delta M_s)^{H^\pm}+ (\Delta
M_s)^{\chi^\pm}+(\Delta M_s)^{\rm DP}|\equiv (\Delta M_s)^{\rm
  SM}|(1+f_s)|
\ee
($\Delta M_s$ is by definition a positive definite quantity).  Here,
$(\Delta M_s)^{\rm SM}$ represents the SM contribution, $(\Delta
M_s)^{H^\pm}$ results from box-diagrams with top and $(H^\pm,H^\pm)$,
$(H^\pm,W^\pm)$ and $(H^\pm,G^\pm)$ exchanges and $(\Delta
M_s)^{\chi^\pm}$ is the contribution of box diagrams with chargino and
squarks.  Finally, $(\Delta M_s)^{\rm DP}$ results from double Higgs
penguin diagrams of fig.~\ref{fig:2pg}.

Explicit expressions for different contributions in terms of the
Wilson coefficients of contributing operators and hadronic matrix
elements can be found in~\cite{BUCHROSL,BUCHROSL02,BUJAUR}.  With
respect to our previous analysis in~\cite{BUCHROSL} we have now
included all resummed large $\tan\beta$ corrections to the relevant
couplings as discussed in the previous section.

In the scenario considered in~\cite{BUCHROSL} and here supersymmetric
particles are heavier than the Higgs particles and the chargino box
contribution $(\Delta M_s)^{\chi^\pm}$ is small.  At large $\tan\beta$
the double penguin contribution $(\Delta M_s)^{\rm DP}$ is the
dominant correction to $(\Delta M_s)^{\rm SM}$ but the charged Higgs
box contribution can also be significant~\cite{BUCHROSL}.  Both
contributions have signs opposite to $(\Delta M_s)^{\rm SM}$.
Consequently for large $\tan\beta$ one finds $(1+f_s)<1$ independently
of the other supersymmetric parameters.  For not too large values of
$\tan\beta\simlt 50$ and of the stop mixing parameter $A_t\simlt
M_{\rm SUSY}$ the contributions $(\Delta M_s)^{\rm DP}$ and $(\Delta
M_s)^{H^+}$ are smaller than $(\Delta M_s)^{\rm SM}$ and one gets
$0<(1+f_s)<1$.  Of interest is also the case $(1+f_s)<0$ corresponding
to a very large negative $(\Delta M_s)^{\rm DP}$ that can be realized
for some special values of supersymmetric parameters - large
$\tan\beta\simgt 50$ and/or $A_t\gg M_{\rm SUSY}$.  We will include
this possibility in our analysis as it has quite different
implications than the case $0<(1+f_s)<1$.

The double penguin diagrams of fig.~\ref{fig:2pg} give ${\cal
  O}(\tan^4\beta)$ correction to $\Delta M_s$.  The leading
contribution comes from the last diagram that contributes to the
Wilson coefficient $C_2^{\rm LR}$ of the operator $Q_2^{\rm
  LR}=(\overline{b_R} s_L) (\overline{b_L} s_R)$.  Using the vertices
of eq.~(\ref{BXRLFIN}) we find~\cite{BUCHROSL02}
\be\label{DMDP1}
(\Delta M_s)^{DP}=\frac{G_F^2 M_W^2}{24\pi^2} M_{B_s} F^2_{B_s}
|V^{\rm eff}_{ts}|^2 P_2^{LR} C_2^{LR}
\ee
where
\begin{eqnarray}\label{C2LRA}
C_2^{\rm LR}\approx -{G_F{\overline m}_b{\overline m}_{d(s)}
{\overline m}_t^4\over\sqrt2\pi^2M^2_W}{\tan^4\beta ~
\epsilon_Y^2 (16\pi^2)^2
\over(1+\tilde\epsilon_3\tan\beta)^2(1+\epsilon_0\tan\beta)^2}
\left[{\sin^2(\alpha-\beta)\over M^2_{H^0}}+
{\cos^2(\alpha-\beta)\over M^2_{h^0}} 
 +{1\over M^2_{A^0}} \right]\nonumber\\
\end{eqnarray}
and $P_2^{LR}\approx 2.5$ includes the short distance NLO QCD
corrections~\cite{BUJAUR,CET0,BUMIUR} and the relevant hadronic matrix
elements~\cite{BECIREVIC}.  Details are given
in~\cite{BUCHROSL,BUCHROSL02,BUJAUR}.  $C_2^{\rm LR}$ in~(\ref{C2LRA})
agrees with the corrected version of~\cite{ISRE}.

For large $\tan\beta$ one has $M_{H^0}\approx M_{A^0}$,
$\cos^2(\alpha-\beta)\approx 0$ and $\sin^2(\alpha-\beta)\approx 1$
and we find
\bea\label{DMDP2}
  &&(\Delta M_s)^{DP}=-12.0/ps \times
\left[\frac{\tan\beta}{50}\right]^4
\left[\frac{P_2^{LR}}{2.50}\right]
  \left[\frac{F_{B_s}}{230\ {\rm MeV}}\right]^2 
  \left[\frac{|V_{ts}|}{0.040}\right]^2\nonumber\\
&&\phantom{aaaaaa}\times
\left[\frac{\overline m_b(\mu_t)}{3.0 {\rm GeV}}\right]
\left[\frac{\overline m_s(\mu_t)}{0.06 {\rm GeV}}\right]
\left[\frac{\overline m_t^4(\mu_t)}{M_W^2 M^2_A}\right] 
\frac{\epsilon_Y^2 (16\pi^2)^2}
{(1+\tilde\epsilon_3\tan\beta)^2(1+\epsilon_0\tan\beta)^2}~.
\eea
We recall that for large $\tan\beta$ the $H^0$ and $A^0$ contributions
to the first two diagrams in fig.~\ref{fig:2pg} cancel each
other~\cite{BAKO,BUCHROSL} and as the contribution of $h^0$ can be
neglected in this limit, the total contributions of these two diagrams
are very small.

\begin{figure}[htbp]
\begin{center}
\begin{picture}(160,80)(0,0)
\ArrowLine(50,40)(20,70)
\ArrowLine(20,10)(50,40)
\Vertex(50,40){5}
\ArrowLine(140,10)(110,40)
\ArrowLine(110,40)(140,70)
\DashLine(50,40)(110,40){3}
\Vertex(110,40){2}
\Text(80,30)[]{\small $h^0$,$H^0$,$A^0$}
\Text(35,70)[]{\small $b_R$}
\Text(40,10)[]{\small $s_L,d_L$}
\Text(125,70)[]{\small $l^-$}
\Text(125,10)[]{\small $l^+$}
\Text(15,40)[]{\small $\tan^2\beta$}
\Text(135,40)[]{\small $\tan\beta$}
\end{picture}
\end{center}
\caption{Dominant diagrams contributing to $B^0_{s,d}\rightarrow l^+l^-$ 
decays at large $\tan\beta$.}
\label{fig:bmumu}
\end{figure}
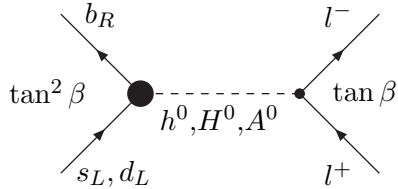

{\bf 2.} At large $\tan\beta$ the branching ratios
$BR(B^0_{s,d}\to\mu^+\mu^-)$ are fully dominated by the diagrams in
fig.~\ref{fig:bmumu}~\cite{BAKO,CHSL,BOEWKRUR,HULIYAZH}.
Following~\cite{BOBUKRUR} we find
\be\label{BRbmumu}
  BR(B_s^0\to\mu^+\mu^-)=2.32\times 10^{-6}
\left[\frac{\tau_{B_s}}{1.5\ ps}\right]
  \left[\frac{F_{B_s}}{230\ {\rm MeV}}\right]^2 
  \left[\frac{|V_{ts}^{\rm eff}|}{0.040}\right]^2
\left[|\tilde c_S|^2+|\tilde c_P|^2\right]~.
\ee
Here $\tilde c_S$ and $\tilde c_P$ are the dimensionless Wilson
coefficients $\tilde c_S=M_{B_s} c_S$ and $\tilde c_P=M_{B_s} c_P$
with $c_S$ and $c_P$ being properly normalized (see~\cite{BOBUKRUR})
Wilson coefficients of the operators
\be\label{ops:bsll:s}
{\Oi}_S = m_b(\overline{b_R} s_L) (\bar{l}l),\quad
{\Oi}_P = m_b (\overline{b_R} s_L) (\bar{l}\gamma_5 l).
\ee

Using the vertices in~(\ref{BXRLFIN}) one finds from the diagrams of
fig.~\ref{fig:bmumu}~\cite{ISRE,BUCHROSL02}
\begin{eqnarray}\label{CS}
c_S\approx
-\frac{m_\mu \overline m_t^2}{4 M^2_W}
\frac{ 16\pi^2 \epsilon_Y ~\tan^3\beta }
{(1+\tilde\epsilon_3\tan\beta)(1+\epsilon_0\tan\beta)}
\left[-{\sin(\alpha-\beta)\cos\alpha\over M^2_{H^0}}
+{\cos(\alpha-\beta)\sin\alpha\over M^2_{h^0}}  \right]~.
\end{eqnarray}

\begin{eqnarray}\label{CP}
c_P\approx
-\frac{m_\mu \overline m_t^2}{4 M^2_W}
\frac{16\pi^2\epsilon_Y ~\tan^3\beta}
{(1+\tilde\epsilon_3\tan\beta)(1+\epsilon_0\tan\beta)}
\left[{1\over M^2_{A^0}}\right]~.
\end{eqnarray}

In the large $\tan\beta$ limit the contribution of $h^0$ to $c_S$ can
be neglected and setting $M^2_{H^0}\approx M^2_{A^0}$ we find
from~(\ref{CS}) and~(\ref{CP}) that $|c_S|=|c_P|$ with $c_P$ given
in~(\ref{CP}).  Consequently
\bea\label{num:BRbmumu}
  BR(B^0_s\to\mu^+\mu^-)&=&3.5\times 10^{-5}
\left[\frac{\tan\beta}{50}\right]^6
\left[\frac{\tau_{B_s}}{1.5\ ps}\right]
  \left[\frac{F_{B_s}}{230\ {\rm MeV}}\right]^2 
  \left[\frac{|V_{ts}^{\rm eff}|}{0.040}\right]^2\nonumber\\
&\times&\frac{\overline m_t^4}{M^4_A} 
\frac{(16\pi^2)^2 \epsilon^2_Y}
{(1+\tilde\epsilon_3\tan\beta)^2(1+\epsilon_0\tan\beta)^2}.    
\eea
This result agrees with \cite{ISRE}. Moreover one has
\be\label{Bdmumu}
\frac{BR(B^0_d\to\mu^+\mu^-)}{BR(B^0_s\to\mu^+\mu^-)} 
=\left[\frac{\tau_{B_d}}{\tau_{B_s}}\right]
  \left[\frac{F_{B_d}} {F_{B_s}}\right]^2 
  \left[\frac{|V_{td}^{\rm eff}|}{|V_{ts}^{\rm eff}|}\right]^2
\left[\frac{M_{B_d}} {M_{B_s}}\right]^5~
\ee
that is, the ratio of the branching fractions can depend on the SUSY
parameters only weakly through $|V_{td}^{\rm eff}/V_{ts}^{\rm eff}|$
which should be consistently determined from the unitarity triangle
analysis~\cite{CHRO,BUCHROSL02}.

The presence of additional $\tan\beta$ dependence in the denominators
of eqs.~(\ref{DMDP2}) and~(\ref{num:BRbmumu}), not included
in~\cite{BUCHROSL} and~\cite{BAKO,CHSL,BOEWKRUR,HULIYAZH}, has been
pointed out in~\cite{ISRE}.  While we confirm these additional
factors, we would like to emphasize that depending on the sign of the
supersymmetric parameter $\mu$ they can suppress $\Delta M_s^{DP}$ and
$BR(B^0_s\to\mu^+\mu^-)$ relative to the estimates in the papers in
question, as stressed in~\cite{ISRE}, but can also provide additional
enhancements.

{\bf 3.} Using~(\ref{DMDP2}) and~(\ref{num:BRbmumu}) we find the
correlation between the neutral Higgs contributions to
$BR(B^0_s\to\mu^+\mu^-)$ and $\Delta M_s^{DP}$ that we have pointed
out in~\cite{BUCHROSL}:
\be\label{CORR}
{BR(B^0_s\to\mu^+\mu^-)}=\kappa ~10^{-6} 
\left[\frac{\tan\beta}{50}\right]^2
\left[\frac{200 {\rm GeV}}{M_{A^0}}\right]^2 
\left[\frac{|\Delta M_s^{DP}|}{2.12/ps}\right]
\ee
where
\be
\kappa=\left[\frac{2.50}{P_2^{LR}}\right]
\left[\frac{3.0 {\rm GeV}}{\overline m_b(\mu_t)}\right]
\left[\frac{0.06 {\rm GeV}}{\overline m_s(\mu_t)}\right]
\left[\frac{\tau_{B_s}}{1.5\ ps}\right]\approx 1~.
\ee
This relation depends sensitively on $M_{A^0}$ and $\tan\beta$ but it
does not depend on $\epsilon_0$ and $\tilde\epsilon_3$.
From~(\ref{Bdmumu}) a similar correlation between
$BR(B^0_d\to\mu^+\mu^-)$ and $\Delta M_s^{DP}$ follows.

In order to understand these results better, let us now assume that
$\Delta M_s$ has been measured and that appropriate supersymmetric
parameters can be found for which the MSSM considered here agrees with
$(\Delta M_s)^{\rm exp}$.  If $0<(1+f_s)<1$ this implies $(\Delta
M_s)^{\rm exp}<(\Delta M_s)^{\rm SM}$.  Then combining~(\ref{DMS0})
and~(\ref{CORR}) we find
\bea
\label{CORRMAIN}
BR(B^0_s\to\mu^+\mu^-)&=&8.5\cdot 10^{-6} \kappa
\left[\frac{\tan\beta}{50}\right]^2 \left[\frac{200 {\rm
      GeV}}{M_{A^0}}\right]^2 \left[\frac{(\Delta M_s)^{\rm
      SM}}{18.0/ps}\right]
\nonumber\\
&\times& \left[1\mp \frac{(\Delta M_s)^{\rm exp}}{(\Delta M_s)^{\rm
      SM}} -\frac{|(\Delta M_s)^{H^\pm}|}{(\Delta M_s)^{\rm SM}}+
  \frac{(\Delta M_s)^{\chi^\pm}}{(\Delta M_s)^{\rm SM}}\right].
\eea 
with ``$\mp$'' corresponding to $0<(1+f_s)<1$ and $(1+f_s)<0$,
respectively.  Using~(\ref{Bdmumu}) analogous expression for
$BR(B^0_d\to\mu^+\mu^-)$ can be found.  In writing~(\ref{CORRMAIN}) we
have taken into account that $(\Delta M_s)^{\rm DP}$ is always
negative and that for large $\tan\beta$ $(\Delta M_s)^{H^\pm}$ is
negative and $(\Delta M_s)^{\chi^\pm}$ is positive.
Formula~(\ref{CORRMAIN}) is valid provided the expression in square
brackets is positive and larger than $10^{-3}$.  Otherwise, other
contributions, in particular those coming from $Z^0$-penguins have to
be taken into account.  In our numerical analysis we take them into
account anyway.

Formula~(\ref{CORRMAIN}) demonstrates very clearly that if $(\Delta
M_s)^{\rm exp}$ will turn out to be close or larger than the SM value,
the order of magnitude enhancements of $BR(B^0_{s,d}\to\mu^+\mu^-)$ in
the scenario of the MSSM considered here with $0<(1+f_s)<1$ will be
excluded.  On the other hand large enhancements of
$BR(B^0_{s,d}\to\mu^+\mu^-)$ are in principle still possible if the
double-penguin contribution is so large that $(1+f_s)<0$ and the "$+$"
sign in~(\ref{CORRMAIN}) applies. For $\tan\beta<50$ obtaining
$(1+f_s)<0$ and the right magnitude of $\Delta M_s$ requires $\mu<0$
so that the couplings~(\ref{BXRLFIN}) are enhanced by the
$\epsilon$-factors in the denominator.  $\mu<0$ is excluded in
particular scenarios like minimal SUGRA, in which the sign of $A_t$ is
fixed and $\mu<0$ does not allow for satisfying the $\bar B\rightarrow
X_s \gamma$ constraint~\cite{CAGANIWA2}, but cannot be excluded in
general.

In order to find $(\Delta M_s)^{\rm exp}/(\Delta M_s)^{\rm SM}$ one
has to deal with the non-perturbative uncertainties contained in the
evaluation of $(\Delta M_s)^{\rm SM}$.  The allowed range for $(\Delta
M_s)^{\rm exp}/(\Delta M_s)^{\rm SM}$ can be obtained by varying all
relevant SM parameters like $m_t$, $V_{ts}$ and
$F_{B_s}\sqrt{B_{B_s}}$.  A conservative scanning of these parameters
performed in~\cite{BUCHROSL} resulted in
\be\label{BOUND}
a \left[\frac{(\Delta M_s)^{\rm exp}}{15/ps}\right]\le 
\frac{(\Delta M_s)^{\rm exp}}{(\Delta M_s)^{\rm SM}}
\le b \left[\frac{(\Delta M_s)^{\rm exp}}{15/ps}\right]
\ee
with $a=0.52$ and $b=1.29$.  It is however clear that the numerical
values of the parameters $a$ and $b$ depend on the error analysis and
the difference $b-a$ should also become smaller as the uncertainties
in the parameters $m_t$, $V_{ts}$ and in particular in
$F_{B_s}\sqrt{B_{B_s}}$ are reduced with time.  For example, the very
recent analysis using the Bayesian approach gives $a=0.71$ and
$b=1.0$~\cite{BUPAST} that correspond to the $95\%$ probability range
$15.1/ps\le(\Delta M_s)^{\rm SM}\le 21.0/ps$.

\begin{figure}[t]
\begin{center}
  \epsfig{file=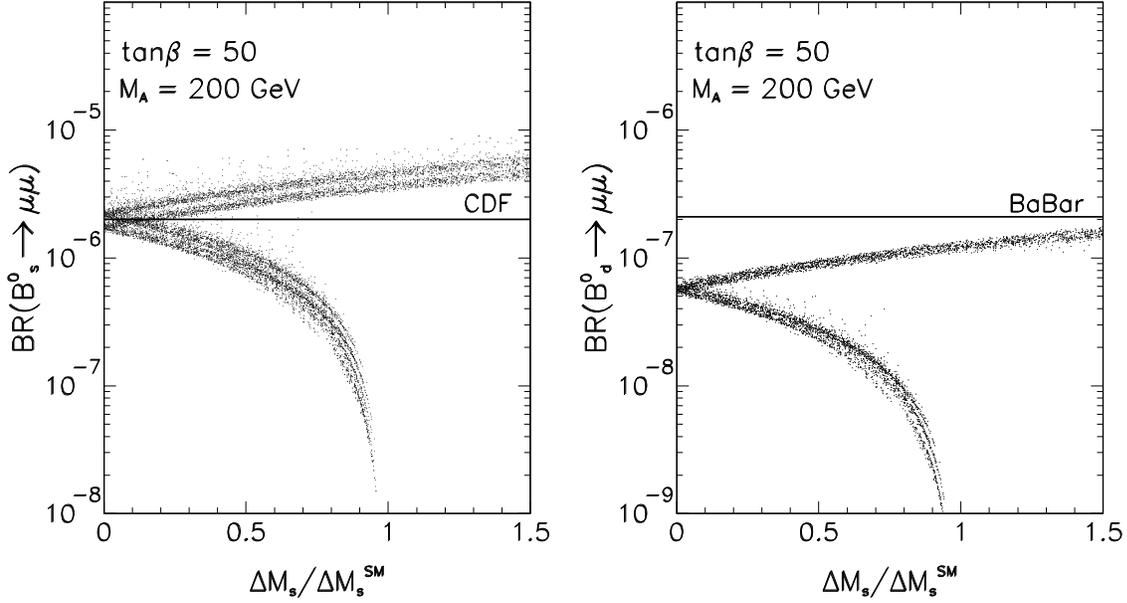,width=\linewidth}
\vspace{-10mm}
\caption{\protect Correlation between $\Delta M_s$ and
  $B^0_{s,d}\ra \mu^+\mu^-$ in the MSSM with flavour violation ruled
  by the CKM matrix.  Lower (upper) branches of points correspond to
  $0<1+f_s<1$ ($1+f_s<0$).  Current experimental bounds: $BR(B^0_s\ra
  \mu^+\mu^-)<2\cdot10^{-6}$ (CDF)~\cite{CDF} and $BR(B^0_d\ra
  \mu^+\mu^-)<2.1\cdot10^{-7}$ (BaBar)~\cite{ALEXAN} are shown by the
  horizontal solid lines.  \label{fig:bcorr}}
\end{center}
\end{figure}

We illustrate the correlations in question in fig.~\ref{fig:bcorr}
where we plot $BR(B^0_{s,d}\to\mu^+\mu^-)$ as functions of $(\Delta
M_s)^{\rm exp}/(\Delta M_s)^{\rm SM}$ for $\tan\beta=50$ and
$M_{A^0}=200$ GeV by scanning the other MSSM parameters with the
restriction that sparticles are heavier than 500 GeV and the $\bar
B\to X_s\gamma$ constraint is satisfied.  For each point in the MSSM
parameter space $V_{td}^{\rm eff}$ is determined by the standard
unitarity triangle analysis \cite{BUCHROSL,CHRO,BUCHROSL02,BUPAST}.
$(\Delta M_d)^{\rm exp}$ and the parameter $\varepsilon_K$ do not
constrain the scan as the Higgs and supersymmetric corrections to
these quantities are small in our scenario \cite{BUCHROSL}.  In the
numerical analysis we have used the formulae from the full
approach~\cite{BUCHROSL,BUCHROSL02} including $SU(2)\times U(1)$
breaking corrections.  Still, the approximate formula~(\ref{CORRMAIN})
describes qualitatively the main features of the correlation.  For
sparticles heavier than 500 GeV the contribution of chargino-stop
boxes to the formula~(\ref{CORRMAIN}) is negligible, $(\Delta
M_s)^{\chi^\pm}/(\Delta M_s)^{\rm SM}\simlt0.03$.  On the other hand,
the contribution of the $H^\pm$ boxes can be substantial, $|(\Delta
M_s)^{H^\pm}|/(\Delta M_s)^{\rm SM}$ can reach $0.65$ due to the
corrections $\epsilon^{HL(R)}$ described in section 3.  This is
contrary to the claim made in ref.  \cite{ISRE} that the
$\epsilon^{HL(R)}$ corrections are not important.  We have checked
that for charginos and stops as light as 150 GeV, $(\Delta
M_s)^{\chi^\pm}/(\Delta M_s)^{\rm SM}\simlt0.2$ whereas $|(\Delta
M_s)^{H^\pm}|/(\Delta M_s)^{\rm SM}$ can reach 0.3.  Also, as follows
from the scan based on the complete calculation, the typical values of
$|(\Delta M_s)^{\rm DP}|$ are smaller for lighter sparticles.

For values of $M_A$ and $\tan\beta$ shown in fig.~\ref{fig:bcorr} all
points corresponding to the rather unlikely scenario with $1+f_s<0$
are eliminated by the combination of the lower limit~(\ref{BOUND}) and
the CDF upper bound $BR(B^0_s\ra
\mu^+\mu^-)<2\times10^{-6}$~\cite{CDF} but this is not the case for
heavier $A^0$ and/or smaller $\tan\beta$ values.  Therefore for such
points we can only use~(\ref{Bdmumu}) to find
\begin{eqnarray}
BR(B^0_d\ra \mu^+\mu^-)<3.6~(3.1)\cdot10^{-8}
\left[{1.15\over F_{B_s}/F_{B_d}}\right]^2
\left[{BR(B^0_s\ra \mu^+\mu^-)^{\rm exp}\over 10^{-6}}\right]
\end{eqnarray}
with the numerical factor corresponding to the analyses
in~\cite{BUCHROSL} and~\cite{BUPAST}, respectively.  With the current
CDF bound one has the upper bound $BR(B^0_d\ra \mu^+\mu^-) <
8~(7)\cdot10^{-8}$ which is still lower than the current BaBar
bound~\cite{ALEXAN}.

For a more likely situation of $0<1+f_s<1$ and $(\Delta M_s)^{\rm
  exp}$ satisfying~(\ref{BOUND}) we get upper bounds on both branching
ratios:
\begin{eqnarray}
&&BR(B^0_s\ra \mu^+\mu^-)\simlt1.2\cdot10^{-6}~(8\cdot10^{-7})
\phantom{aaa}{\rm for}\phantom{aa} 
  a=0.52~(0.71) ,\nonumber\\
&&BR(B^0_d\ra \mu^+\mu^-)\simlt3\cdot10^{-8}~(2\cdot10^{-8})
\phantom{aaa}{\rm for}\phantom{aa} a=0.52~(0.71) .   
\label{eqn:BOUNDF}
\end{eqnarray}
where the two values for the parameter $a$ correspond to the analyses
in~\cite{BUCHROSL} and~\cite{BUPAST}, respectively. This should be
compared with the SM values that are in the ballpark of $3\cdot
10^{-9}$ and $1\cdot 10^{-10}$, respectively. On the basis of our
discussion of the contribution $(\Delta M_s)^{\chi^\pm}$, we would
like to emphasize that the upper limits on
$BR(B^0_{s,d}\ra\mu^+\mu^-)$ obtained here for heavy sparticle
spectrum cannot be significantly altered by lowering the sparticle
masses.

\section{Summary}

In this letter we have analyzed $\Delta M_s$ and $BR(B^0_{s,d}\ra
\mu^+\mu^-)$ in the MSSM with the CKM matrix as the only source of
flavour and CP violation.  By considering heavy sparticle spectrum we
have quantified the tight correlation between these quantities that
exists for large values of $\tan\beta$.  Our analysis shows that the
neglect of this correlation in the analyses of $BR(B^0_{s,d}\ra
\mu^+\mu^-)$ at large $\tan\beta$ as done in the previous
literature~\cite{BAKO,CHSL,BOEWKRUR,HULIYAZH,ISRE,DEDRNI} is not
justified.  The correlation in question leads to interesting upper
bounds on $BR(B^0_s\ra \mu^+\mu^-)$ and $BR(B^0_d\ra \mu^+\mu^-)$ not
considered sofar in the literature.  In the most likely scenario with
$0<(1+f_s)<1$ the upper bounds are becoming very strong when the ratio
$(\Delta M_s)^{\rm exp}/ (\Delta M_s)^{\rm SM}$ approaches unity.  For
$(\Delta M_s)^{\rm exp} \ge (\Delta M_s)^{\rm SM}$ substantial
enhancements of $BR(B^0_{s,d}\ra \mu^+\mu^-)$ with respect to the
values obtained in the SM are not possible within the MSSM scenario
considered here.  Therefore finding experimentally $BR(B^0_d\ra
\mu^+\mu^-)$ above $3\cdot10^{-8}$, that is one order of magnitude
below the current limit, would be a strong signal of new sources of
flavour violation~\cite{CHRO}.

As the upper bounds on $BR(B^0_{s,d}\ra \mu^+\mu^-)$ discussed here
are sensitive functions of the ratio $(\Delta M_s)^{\rm exp}/ (\Delta
M_s)^{\rm SM}$, their quantitative usefulness will depend on the value
of $(\Delta M_s)^{\rm exp}$ and on the accuracy with which $(\Delta
M_s)^{\rm SM}$ can be calculated.  In this respect the present efforts
of experimentalists to measure $BR(B^0_{s,d}\ra \mu^+\mu^-)$ and
$\Delta M_s$ and of theorists to calculate $F_{B_{d,s}}$ and the
parameters $B_{d,s}$ appear even more important than until now.

\section*{Acknowledgments}

A.J.B. would like to thank A. Dedes, F. Kr\"uger and J. Urban for
their interest and discussions.  The work of A.J.B. and J.R. has been
supported in part by the German Bundesministerium f\"ur Bildung und
Forschung under the contract 05HT1WOA3 and the DFG Project Bu.
706/1-1.  The work of P.H.Ch. has been partly supported by the Polish
State Committee for Scientific Research grant 5~P03B~119~20 for
2001-2002 and by the EC Contract HPRN-CT-2000-00148 for years
2000-2004.

\end{document}